\PassOptionsToPackage{table,dvipsnames}{xcolor}
\PassOptionsToPackage{draft}{hyperref}
 \documentclass[sigconf, table, natbib=true]{acmart}

\usepackage[utf8]{inputenc}
\makeatletter%
\@ifclassloaded{acmart}{
  \makeatletter
  \def\mdseries@tt{m}
  \makeatother
}{}
\makeatother

\usepackage{graphicx}
\usepackage[utf8]{inputenc} 
\usepackage[T1]{fontenc}    
\usepackage{booktabs}       
\usepackage{amsfonts}       
\usepackage{nicefrac}       
\usepackage{microtype}      
\usepackage{amsmath}
\usepackage{multirow} 
\usepackage{courier}

\usepackage{adjustbox} 
\usepackage{tabularx}

\newtheorem{theorem}{Theorem}

\newtheorem{definition}{Definition}

\newcommand{\textcite}[1]{\citet{#1}}

\begin{document}

\acmYear{} 
\acmPrice{} 
\acmISBN{} 
\acmDOI{} 
\setcopyright{none} 
\acmConference[AdvML'19: Workshop on Adversarial Learning Methods for Machine Learning and Data Mining at KDD]{AdvML'19: Workshop on Adversarial Learning Methods for Machine Learning and Data Mining at KDD}{}{}

\title
{Connecting Lyapunov Control Theory to Adversarial Attacks}

%

 \author{Arash Rahnama}
   \email{Rahnama\_Arash@bah.com}
   \affiliation{%
     \institution{Booz Allen Hamilton}
     }
     
 \author{Andre T. Nguyen}
   \email{Nguyen\_Andre@bah.com}
   \affiliation{%
     \institution{Booz Allen Hamilton}
     }
     
 \author{Edward Raff}
   \email{Raff\_Edward@bah.com}
   \affiliation{%
     \institution{Booz Allen Hamilton}
   }

\begin{abstract}
Significant work is being done to develop the math and tools necessary to build provable defenses, or at least bounds, against adversarial attacks of neural networks. In this work, we argue that tools from control theory could be leveraged to aid in defending against such attacks. We do this by example, building a provable defense against a weaker adversary. This is done so we can focus on the mechanisms of control theory, and illuminate its intrinsic value. 
\end{abstract}

\settopmatter{printfolios=true} 

\maketitle

\section{Introduction}
Adversarial Machine Learning has been a research area for over a decade \cite{Lowd:2005:AL:1081870.1081950}, but has only recently gained increased attention due to the successful application of adversarial attacks to deep learning networks \cite{Goodfellow2015}. If we define the DNN as a model $f(\cdot)$, which produces an output $y$ given some input $x$, then we are interested in the types of adversarial attacks which can perturb $x$ given some change $\Delta$ such that the model produces an incorrect decision (i.e., $f(x) \neq f(x + \Delta)$). Typically, it is assumed that the attack parameter $\Delta$ is bounded by some $L_p$ norm, such that $\|\Delta\|_p \leq \epsilon$. Adversarial attacks have been successful in degrading the performance of DNNs across many domains and algorithms even with very small values of $\epsilon$ \cite{Biggio2017}. For instance, it has been shown that by altering only one pixel in the input image, the standard convolutional neural network can be fooled into making the wrong decision (i.e., $\|\Delta\|_0 \leq 1$) \cite{Su2017}. This points to the innate vulnerability of DNNs and its negative impacts on the public trust in reliability and safety of machine learning systems. As the use of DNNs in safety-critical environments such as autonomous vehicles \cite{Eykholt2018} and medicine \cite{Fredrikson2014} increases, the need for designing provable defensive solutions against adversarial attacks also grows. This also motivates our interest in designing robust deep learning models.

With an increased interest from the community, many attempts to produce heuristic defensive designs have been proposed \cite{Yuan2017,Kannan2018,Li2017a,Guo2018,Xie2018,Samangouei2018}. Yet when carefully evaluated, it has been routinely found that these defensive approaches are not effective against their intended adversary \cite{Athalye2018,Athalye2018a}. Another venue of research has focused on building provably robust defensive designs against adversarial attacks. These have so far focused on the careful application of more sophisticated optimization techniques to prove that everything within an $L_p$ ball of the training data will produce the same output \cite{Abbasi2018,Gowal2018,Wong2018,Dvijotham2018,pmlr-v80-wong18a}. While the motivation behind these optimization based approaches is intuitive, the model must be robust if the response is consistent, none of these methods can yet scale to large datasets. Given the shocking nature by which most defensive designs have been easily defeated \cite{Carlini:2017:AEE:3128572.3140444}, many have begun the work to build a new theory which can help explain and resolve these issues. This work has been done "from the ground up," and attempts to build new mathematical tools and results to understand the problem. Early works provided grounding to the intuitive connection between model's accuracy, the dimensionality of the feature space, and the model's susceptibility to attack \cite{Gilmer2018}. \textcite{Wang2017} developed foundations to compare a trained model $f^t$ to an oracle $f^o$, and provided a connection between extraneous features and susceptibility to attack. \textcite{Demontis} showed connections between the norm of the input gradients and the ability to \textit{transfer} attacks against one model to a second unknown model. 

In this work, we make an important connection between the field of control theory and the design of robust DNNs. 
In \autoref{sec:main} we show how to use
Lyapunov theory of stability 
to model neural networks as a 
dynamical nonlinear system, and 
bound the perturbation's effect against a simple adversary for all possible inputs. 
Breifly, we will discuss related work in \autoref{sec:related}.
An abridged review of the needed control theory, and empirical validation of our theory, are  available in the appendix. 
\section{Main Results} \label{sec:main}
Using the above results from the field of control theory, we show how to develop a regularization technique that provides provable bounds for DNN with $N$ layers. The primary result is given in \eqref{eq:final_bound_general}, where $\Delta^{(1)}$ is the change in the input of the DNN, which is constrained such that $\Delta^{(1)}_i = \Delta^{(1)}_j$, $\forall i, j$. $\forall l  \in [2, N], \Delta^{(l)}$ is the resulting perturbations to the activation of each proceeding layer of the network. The constants $\epsilon$, $\beta$,  can be chosen almost arbitrarily, so long as the denominator remains positive. 
A third parameter $\rho$ must satisfy $\rho>{\cos\left(\pi/({N+1})\right)^{N+1}} \cdot \left(\prod_{l=1}^{N} \nu_l\right)^{-1}$. This shows the deviations of the network's final activation ($\Delta^{(N)}$) is bounded by a ratio of the input perturbation size ($\Delta^{(1)}$) for \textit{all} possible inputs. 

\begin{equation} \label{eq:final_bound_general}
\|\Delta^{(N)}\|^2_2
\leq  \frac{\beta \|\Delta^{(1)}\|_2^2}{2 \cdot \left(\epsilon - \rho  - \frac{1}{2\beta}\right) } 
\end{equation}

Our proof strategy begins with treating each layer of the DNN as a nonlinear dynamical system. For each layer, we show the conditions under which the layer obtains the Incrementally Input Feed-Forward Passive (IIFP) properties (see \cite{zames1966input}, or Definition \ref{IIFP} ).
The control theory view allows us to consider each layer independently in our analysis, where the input to one layer is the output of the previous. 
Using a sequence of IIFP layers, we then show that their sequential combination, under certain conditions, will maintain the Incrementally Output Feedback Passive (IOFP) property (see \cite{zames1966input}, or Definition \ref{IOFP}). Having the IOFP property allows us to derive the global bound given in Equation \ref{eq:final_bound_general}, producing a DNN which is provably robust against an adversary that can alter the input by any constant factor. This result can be used both to understand robustness for classification problems, and the less studied regression case \cite{Nguyen2019}.

\subsection{Proving Robustness for the Cascade }
Our aim is to find a relationship between the distortions introduced by adversarial examples and robustness in DNNs. Here, we characterize a measure of robustness which can be used to certify a minimum performance index against adversarial attacks on a neural network. Given a DNN, we are interested in characterizing the (local) robustness of an arbitrary natural example $u$ by ensuring that all of its neighborhood has the same inference outcome. The neighborhood of $u$ is characterized by an $L_2$ ball centered at $u$. Geometrically speaking, the minimum distance of a misclassified nearby example to $u$ is the least adversary strength required to alter the target model's prediction, which is also the largest possible robustness certificate for $u$. We aim to utilize the IIFP and IOFP properties of the activation function and their relationship with Lyapunov stability properties of nonlinear systems to find a robustness measure \cite{zames1966input}. The definition of the injected perturbations by the adversary is as follows, 
\begin{definition}
	Consider the input $u$ to the layer of size $n$, $u\in \mathbb{R}^n$. The perturbed input signal is  $u+\Delta$ where $\Delta \in \mathbb{R}^n$ is the attack vector with all positive or all negative entries of the same size. The perturbed input vector $u$ is within an $\epsilon$-bounded $L_p$-ball centered at $u$ i.e., $u \in B_p(u,\epsilon)$, where $B_p(u,\epsilon):=\{u+\Delta|~\|u+\Delta-u\|_p=\|\Delta\|_p\leq\epsilon\}$.
 \label{attack}
\end{definition}
Here, we consider the constant variations $\Delta^{(k)}$, i.e., $\Delta^{(k)}_i = \Delta^{(k)}_j$, $\forall i, j$, which are injected by the adversary into the initial input or the signals traveling from a hidden layer to another. A system is defined as a layer inside the DNN which accepts an input of size $n_{l-1}$ (output of the previous layer) and produces an output of size $n_{l}$ (what is produced after the activationn transformation). We suggest that DNN's parameters should be trained so that the output variations are small for small variations in input $u$. We treat each layer of the DNN as a nonlinear dynamical system. We show the conditions under which a layeris IIFP from its input to its output. Then, we prove that the interconnection of IIFP layers under certain conditions is IOFP with a negative $\rho$ and as a result find a bounded stable and robust relationship between the input and output of the entire DNN i.e., show that bounded changes applied to the input produce bounded changes in the output which are upper-bounded by the changes in the input. 
\begin{theorem}
	 Consider the cascade interconnection of nonlinear systems 
    $H_{1, \ldots, N}$
	 where $N>2$, if each sub-system $H_i$ for $i=1,...,N$ is instantaneously Incrementally Input Feed-Forward Passive (IIFP) with a storage function $V_i$ and $\nu_i>0$ such that, 
	 \begin{equation*}
	\dot{V_i}=\omega(u_{i2}-u_{i1},y_{i2}-y_{i1})=(u_{i2}-u_{i1})^T(y_{i2}-y_{i1})-\nu_i (u_{i2}-u_{i1})^T(u_{i2}-u_{i1}),
	 \end{equation*}
	 
	 then there exists a positive $\rho$, where,
	 $\rho>{\cos\left(\frac{\pi}{N+1}\right)^{N+1}} \cdot \left(\prod_{l=1}^{N} \nu_l\right)^{-1}$.
	 
	 for which the entire cascade interconnection of nonlinear systems admits a storage function of the form,
	 
	 \begin{equation} \label{Store}
	 V = \sum_{i=1}^{N} d_i V_i,
	 \end{equation}
	 
	 satisfying 
	 
	 \begin{equation} \label{All_Store}
	 \dot{V} \leq -\epsilon y^Ty+\rho y_N^Ty_N +u_1y_N,
	 \end{equation}

	for some $\epsilon>0$, where $y = [y_{12}-y_{11}^T,y_{22}-y_{21}^T, \dots, y_{N2}-y_{N1}^T]^T$, and $u_{i2}-u_{i1}$ and $y_{i2}-y_{i1}$ are the incremental inputs and outputs of system $i$.
 \label{Th_Stabb} 
 \end{theorem}
\bf{Proof:} \normalfont The proof is to show that the cascade interconnection of the systems is IOFP with the passivity index $-\rho$. We need to show that the storage function given in (\ref{Store}) satisfies \eqref{All_Store},
where $\dot{V_i}\leq(u_{i2}-u_{i1})^T(y_{i2}-y_{i1})-\nu_i (u_{i2}-u_{i1})^T(u_{i2}-u_{i1})$, and $u_i=u_{i2}-u_{i1}$ is the incremental difference between any two input signals to the layer (system) $i$, and $y_i=y_{i2}-y_{i1}$ is the incremental difference between their respective outputs for the layer (system) $i$. $u_1=u_{12}-u_{11}$ is the incremental input to the first layer (system), and $y_N=y_{N2}-y_{N1}$ is the final incremental output of the cascade interconnection and $y$ is a vector of size $N*1$ with incremental output entries $y_i$ where $i=1,...,N$. The relationship given in (\ref{All_Store}) holds if,
\begin{equation} \label{Net_Store}
\sum_{i=1}^{N} [d_i (u_i^Ty_i-\nu_i u_i^Tu_i)] -\rho y_N^Ty_N -u_1y_N\leq -\epsilon y^Ty.
\end{equation}
We can define,
\begin{equation*} A = 
\begin{bmatrix}
 -\nu_1& 0 & \dots & 0 & -\frac{1}{\rho}\\
1& -\nu_2 & \ddots & & 0  \\
0 &1 & -\nu_3  & \ddots & \vdots \\
\vdots & \ddots & \ddots & \ddots & 0 \\
0 & \dots & 0 & 1 & -1
\end{bmatrix}, \rho_i > 0, \nu_i > 0,
\end{equation*}
and $D=diag\{\rho,d_1,d_2,\dots,d_n\}$. Then it can be seen that the left hand-side of (\ref{Net_Store}) is equal to
\begin{equation*}
[u^T y^T] DA \begin{bmatrix}
u \\ y
\end{bmatrix} = \frac{1}{2} [u^T y^T] [DA+A^TD]  \begin{bmatrix}
u \\ y
\end{bmatrix} 
\end{equation*}
where $y=[(y_{12}-y_{11})^T,\dots,(y_{N2}-y_{N1})^T]^T$. According to Theorem \ref{Th_Sec} if,
$\rho>{\cos\left(\frac{\pi}{N+1}\right)^{N+1}} \cdot \left(\prod_{l=1}^{N} \nu_l\right)^{-1}$,
then there exists a diagonal matrix $D>0$ such that $DA+A^TD<0$. Hence, it can be shown that the left size of the equation given in (\ref{Net_Store}) is negative, i.e.,
$
[u^T y^T] DA \begin{bmatrix}
u \\ y
\end{bmatrix}< -\epsilon y^Ty
$
and thus,
$
\dot{V} \leq -\epsilon y^Ty +\rho y_N^Ty_N +u_1y_N.
$


We have formulated the storage function, given in (\ref{All_Store}), for the cascade of systems 
$H_{1 \ldots N}$.
As mentioned before, each system 
$H_i$
represents a layer inside the DNN and the entire cascade interconnection represents the entire DNN. Now, we can characterize a relationship between the incremental inputs fed into the first layer of DNN and the respective incremental outputs at the final layer of the DNN. This means that we can effectively characterize the changes caused by adversarial attacks by quantifying their effects on the output of DNN. We can prove an upper-bound for the changes occurring at the output layer of the DNN given the respective input differences fed into the network. Needless to say, these input differences are caused by adversarial attacks. Consequently, we characterize a measure of robustness for the entire DNN. As shown in the next corollary, if the loss function is designed such that it is encouraged for each hidden layer of the DNN to behave as an IIFP nonlinear system then the changes in the output caused by the attack vector $\Delta^{(i)}$ injected by the adversary into the input of the layer $i$ are bounded and limited by the changes in the input signal itself (the norm of the attack parameter). Correspondingly, the adversary's ability to change the output behavior is limited to the use of larger attack parameters which in return are easier to detect.

\begin{corollary}
Consider a cascade of $H_{1, \ldots N}$ nonlinear systems organized as feed-forward layers of a neural network, with $N> 2$ layers. 
If each sub-system $H_i$ is instantaneously Incrementally Input Feed-Forward Passive (IIFP) with a positive incremental input passivity index $\nu_i>0$, then the entire cascade of systems is instantaneously Incrementally Output Feedback Passive (IOFP) with the passivity index $-\rho$ and the storage function $w(x,y) = -\epsilon y^Ty +\rho y_N^Ty_N +u_1y_N$ where $\rho$ meets the condition,
$\rho> \cos\left(\pi/(N+1)\right)^{N+1} \cdot \left( \prod_{i=1}^N \nu_i \right)^{-1}$.
One can show that the variations in the final output of the entire network ($\Delta^{(N)}$) are upper-bounded (limited) by the variations in the input signal ($\Delta^{(1)}$) through the following relation,
\begin{align*}
 \left(\epsilon - \rho  - \frac{1}{2\beta}\right)\|\Delta^{(N)}\|^2_2
\leq  \frac{\beta \|\Delta^{(1)}\|_2^2}{2},
\end{align*}
or further as a tighter bound, for all the output variations for all the layers, we can show the following bound, where $\epsilon>0, \beta>0$ and $[\epsilon - \rho  - \frac{1}{2\beta}]>0$.
\begin{align*}
\epsilon \cdot \sum_{i=2}^{N-1} \|\Delta^{(i)}\|_2^2  + \left(\epsilon - \rho  - \frac{1}{2\beta} \right) \|\Delta^{(N)}\|_2^2
\leq \frac{\beta \|\Delta^{(1)}\|_2^2}{2}
\end{align*}

\end{corollary}

\bf{Proof:}\normalfont~Given Theorem \ref{Th_Stabb}, and the following definitions, $\Delta^{(N)} =y_N=y_{N2}-y_{N1}$, $\Delta^{(i)} =y_i=y_{i2}-y_{i1}$ and $\Delta^{(1)} = u = u_1 + \Delta^{(1)} - u_1$, we have,
\begin{align*}
 &0 \leq -\epsilon y^Ty +\rho y_N^Ty_N +u_1y_N
\\& \leq -\epsilon \cdot \sum_{i=1}^{N} \|\Delta^{(i)}\|_2^2 + \rho \|\Delta^{(N)}\|_2^2
\\& \leq -\epsilon \cdot \sum_{i=1}^{N} \|\Delta^{(i)}\|_2^2 + \rho \|\Delta^{(N)}\|_2^2
 - \left(\frac{\sqrt{\beta} u}{\sqrt{2}}-\frac{\Delta^{(N)}}{\sqrt{2\beta}}\right)^2 + \frac{\beta \|u\|_2^2}{2}+\frac{\|\Delta^{(N)}\|_2^2}{2\beta}
\\& \leq -\epsilon \|\Delta^{(N)}\|_2^2 + \rho \|\Delta^{(N)}\|_2^2 + \frac{\beta \|u\|_2^2}{2}+\frac{\|\Delta^{(N)}\|_2^2}{2\beta}
\end{align*}
where $\epsilon>0, \beta>0$ are design parameters and $\rho>\frac{\cos(\frac{\pi}{N+1})^{N+1}}{\nu_1\times\nu_2\dots\times\nu_N}$.
Finally if we move the appropriate terms to the left side of the above inequalities we have,
\begin{align*}
&\left(\epsilon - \rho  - \frac{1}{2\beta}\right)\|\Delta^{(N)}\|_2^2
 \leq \frac{\beta u^2}{2} =  \frac{\beta \|\Delta^{(1)}\|_2^2}{2}.
\end{align*}
Or further as a tighter bound we can have,
\begin{align*}
& \epsilon \cdot \sum_{i=2}^{N-1} \|\Delta^{(i)}\|_2^2 + \left(\epsilon - \rho  - \frac{1}{2\beta}\right)\|\Delta^{(N)}\|_2^2
\leq \frac{\beta \|\Delta^{(1)}\|_2^2}{2}
\end{align*}
where  $\left(\epsilon - \rho  - \frac{1}{2\beta}\right)>0$ should hold. 


\subsection{Proving Bounds Against Perturbations }
A DNN can be represented as a cascade of systems. One can model the DNN as $y_l = f_l(W_l u_{l-1} + b_l)$ for $l = 1,...,N$ for some $N>2$, where $u_{l-1} \in \mathbb{R}^{n_{l-1}}$ is the input feature of the $l$-th layer, $f_l : \mathbb{R}^{n_{l-1}} \rightarrow \mathbb{R}^{n_l}$ is a (non-linear) activation function, and $W_l \in \mathbb{R}^{n_l \times n_{l-1}}$ and $b_l \in \mathbb{R}^{n_l}$ are respectively the layer-wise weight matrix and bias vector applied to the flow of information from the layer $l-1$ to $l$. $n_{l-1}$ and $n_{l}$ represent the number of neurons in layers $l-1$ and $l$. For a set of parameters, $\Theta = \{W_l, b_l\}^{N}_{l=1}$, we denote the function representing the entire DNN as 
$f_\Theta(u^{(1)}) = u^{(N)}$ where $f_\Theta : \mathbb{R}^{n_1} \rightarrow \mathbb{R}^{n_N}$. Given the training data, $(u_i , y_i )^{K}_{i=1}$, where $u_i \in \mathbb{R}^{n_1}$ and $y_i \in \mathbb{R}^{n_N}$, the loss function is 
defined as $\frac{1}{K} L(f_{\Theta}(u_i), y_i)$, where $L$ is usually selected to be cross-entropy or the squared $L_2$-distance for classification and regression tasks, respectively. The model parameter to be learned is $\Theta$. We consider how we can obtain a model that is insensitive to the perturbation of the input. The goal is to obtain a model, $\Theta$, such that the $L_2$-norm of the incremental change $f(u_1 + \Delta) - f(u_1)$ is small, where $u_1 \in \mathbb{R}^{n_1}$ is an arbitrary vector and $\Delta \in \mathbb{R}^{n_1}$ is a perturbation vector with a small $L_2$-norm. Most DNNs exhibit nonlinearity only due to the activation functions, such as ReLU, maxout and maxpooling. In such cases, function $f_\Theta$ is a piece-wise linear function. Hence, if we consider a small neighborhood of $u$, we can regard $f_\Theta$ as a linear function. In other words, we can represent it by an affine map, $u \rightarrow  W_{\Theta, u} u + b_{\Theta, u} $, using a matrix, $W_{\Theta,u} \in \mathbb{R}^{n_1\times n_L}$, and a vector, $b_{\Theta,u} \in \mathbb{R}^{n_L}$, which depend on $\Theta$ and $u$. It is important to note that because of Theorem \ref{Th_Sec}, the number of layers in the DNN under consideration should be larger than $2$ ($N>2$). This does not limit our results as any DNN with a smaller number of layers will only consist of an input and an output layer. 

We suggest that model parameter $\Theta$ should be trained so that the output variations are small for small variations in input $u$. To further investigate the property of $W_{\Theta,u}$, we assume that each activation function, $f_l$ is a modified version of element-wise ReLU called the Leaky ReLU: 
$f_l(u_{l-1}) = \max(y_l, a \cdot y_l )$ 
where $y_l = W_l u_{l-1} + b_l$, and 
$0<a<1$. It follows that, to bound the variations in the output of the neural network by the variations in the input, it suffices to bound these variations for each $l \in \{1, ..., L\}$. Here, we consider that the variations $\Delta$ are injected by the adversary into the initial input or the signal traveling from a hidden layer to another. This motivates us to consider a new form of regularization scheme, which is described in the following. As mentioned before, a system is defined as a layer inside the DNN which accepts an input of size $n_{l-1}$ (output of the previous layer) and produces an output of size $n_{l}$ (what is produced after the Leaky ReLu transformation). 

The transformations between the two layers $l-1$ and $l$ can be divided into two sub-transformations which respectively represent the set of row operations done on the input signal that produce positive or negative outputs on the other side of the Leaky ReLU activation function. The positive transformation includes the rows $[f_l(u_{l-1})]_{n^+} = [W^{n^+}_l u^{n^+}_{l-1} + b^{n^+}_l]\geq0$ and the negative transformation includes the rows $[f_l(u_{l-1})]_{n^-}= [(W^{n^-}_l u^{n^-}_{l-1} + b^{n^-}_l)]<0$ where ${n^+}+{n^-}=n_l$. Below, if the consecutive layers $l$ and $l-1$ are of different sizes, the appropriate matrices are padded with zeros. These changes do not affect our results and are only done for mathematical tractability. If $n_{l-1} > n_l$, then $W_l$ is padded with rows of zero i.e. $W \in \mathbb{R}^{n_{l-1} \times n_{l-1}}$ and the identity matrix $I$ has the dimensions of $I \in \mathbb{R}^{n_{l-1} \times n_{l-1}}$. If $n_{l-1}<n_l$, then $W_l$ is padded with columns of zero i.e. $W_l \in \mathbb{R}^{n_l \times n_l}$, the identity matrix initially of size $I \in \mathbb{R}^{n_{l-1} \times n_{l-1}}$ is padded with rows and columns of zero to produce $I \in \mathbb{R}^{n_l \times n_l}$ and the $\Delta^{(l-1)}$ vector initially of size $\Delta^{(l-1)} \in \mathbb{R}^{1 \times n_{l-1}}$ is padded with rows of zero to produce $\Delta^{(l-1)}\in \mathbb{R}^{1 \times n_l}$. We need to first show the conditions under which the non-linear transformations inside the DNN are IIFP each with a positive input passivity index $\nu>0$. As a result, the input passivity index for the layer $l$, $\nu_l$ is a positive design parameter representing the extend to which, we want to encourage this behavior in the sub-layer $l$. These parameters will re-appear in the loss function for the entire system to encourage this behavior on the network level. 

Given the Definition \ref{IIFP}, and the fact that what happens at the output level constitutes a linear transformation, we have,
\begin{align} \label{relation1}
&\omega(u_{l-1} + \Delta^{(l-1)} - u_{l-1}, f(W_l [u_{l-1} + \Delta^{(l-1)}] + b_l)-f(W_l u_{l-1} + b_l)) \nonumber\\&=(\Delta^{(l-1)})^T\Lambda W_l\Delta^{(l-1)}- (\Delta^{(l-1)})^T\nu_l I\Delta^{(l-1)} \nonumber\\& 
=(\Delta^{(l-1)})^T\Lambda W_l\Delta^{(l-1)}-n_{l-1}\nu_l(\Delta^{(l-1)})^T\Delta^{(l-1)}
\end{align}
for some positive $\nu_l \in \mathbb{R}$. $\Lambda \in \mathbb{R}^{max(n_l,n_{l-1})\times max(n_l,n_{l-1})}$  is a diagonal matrix defined as follows:
the first $n^+$ diagonal entries are equal to $1$, the next $n^-$ diagonal entries are equal to $a$ and the rest of the $max(n_l,n_{l-1})-n^+-n^-$ diagonal entries are zeros. The relationship given in (\ref{relation1}) can be further simplified to have,
\begin{align*}
&\omega(u_{l-1} + \Delta^{(l-1)} - u_{l-1}, f(W_l [u_{l-1} + \Delta^{(l-1)}] + b_l)-f(W_l u_{l-1} + b_l)) \nonumber\\&=(\Delta^{(l-1)})^T\Lambda W_l\Delta^{(l-1)}-n_{l-1}\nu_l(\Delta^{(l-1)})^T\Delta^{(l-1)} \nonumber\\&= [\sum_{i=1}^{n^+}\sum_{j=1}^{n_{l-1}} w^l_{ij}+a\sum_{i=1}^{n^-}\sum_{j=1}^{n_{l-1}} w^l_{ij}](\Delta^{(l-1)})^T\Delta^{(l-1)}-n_{l-1}\nu_l(\Delta^{(l-1)})^T\Delta^{(l-1)}\nonumber\\&\geq a[\sum_{i=1}^{n^+}\sum_{j=1}^{n^{l-1}} w^l_{ij}+\sum_{i=1}^{n^-}\sum_{j=1}^{n^{l-1}} w^l_{ij}](\Delta^{(l-1)})^T\Delta^{(l-1)}-n_{l-1}\nu_l(\Delta^{(l-1)})^T\Delta^{(l-1)}\\& \geq 0
\end{align*}
The above relation holds and as a result the above transformations are IIFP, if the summation of the weights (entries of $W$) is greater than $n_{l-1}\times\frac{\nu_l}{a}$, i.e.,  $(\sum_{i=1}^{n_l} \sum_{j=1}^{n_{l-1}} w^l_{ij})>n_{l-1}\times \frac{\nu_l}{a}$.

As a result a regularization scheme that would encourage the layers to behave as IIFP nonlinear systems should encourage the following relation for each layer,
\begin{align} \label{key}
\left(\sum_{i=1}^{n_l} \sum_{j=1}^{n_{l-1}} w^l_{ij}\right)>n_{l-1}\times \frac{\nu_l}{a}~For~l=1,...,N.
\end{align}
where $n_{l-1}$ is the number of neurons in the previous layer, $n_{l}$ is the number of neurons in the next layer and $N$ is the number of hidden layers. The above regularization rule on the weights is happening at the layer level independent of other layers, unlike Ridge or LASSO regularization rules. A simple regularization rule added to the loss function that encourages the behavior given in (\ref{key}) will maintain the IIFP property for each layer and the IOFP property with a negative $\rho$ for the entire DNN as defined in Theorem \ref{Th_Stabb}. Finally, the variations in the final output of the entire network ($\Delta^{(N)}$) is upper-bounded (limited) by the variations in the input signal ($\Delta^{(1)}$) through the following relation,
\begin{align*}
\left(\epsilon - \rho  - \frac{1}{2\beta}\right)\|\Delta^{(N)}\|^2_2
\leq  \frac{\beta \|\Delta^{(1)}\|_2^2}{2} 
\end{align*}
Or similarly as a tighter bound, for all the output variations at all the layers we have \eqref{eq:tight_bound}, where  $(\epsilon - \rho  - \frac{1}{2\beta})>0$. We point out that the view of DNNs as a non-linear system $H_{i, \ldots, N}$ does not depend on any special properties, or even recognition that the network has a final layer. As such, \textit{the bounds apply to all layers simultaneously, bounding an attack initiated at any individual layer, as well as the response of any hidden layer}. This type of result has never been previously shown, and comes for free with control theory. 
\begin{equation} \label{eq:tight_bound}
\epsilon \cdot \sum_{i=2}^{N-1} \|\Delta^{(i)}\|_2^2 + \left(\epsilon - \rho  - \frac{1}{2\beta}\right)\|\Delta^{(N)}\|_2^2
\leq \frac{\beta \|\Delta^{(1)}\|_2^2}{2}
\end{equation}

\section{Related Work} \label{sec:related}

While we are not aware of any prior work that has shown the direct applicability of control theory to adversarial attacks, we make note of two types of connections to prior work. 

First, \textcite{Zantedeschi:2017:EDA:3128572.3140449} developed a bounded (or "clamped") version of the ReLU activation function as a method of bounding the perturbation of the network independent of the learned network's weights. By doing so they show $\forall u_1, \Delta^1,~  \|\Delta^N\|^2_2 \leq  M \|\Delta^1\|_2^2$ where $M=\prod_{j=1}^{N} M_j$ is the product of the Lipschitz constants of each layer. This can be seen as an special case of our results where the parameters are selected such that $\frac{\beta}{2(\epsilon - \rho  - \frac{1}{2\beta})} = M$. 

Second, we note as an example the valuable work of \textcite{NIPS2018_7742}, who developed bounds on a network's response with a variety of activation functions. In the parlance of control theory, their work shows similar bounded conic behavior as what we have performed in this work. In contrast to their work, our approach leverages control theory to define the behavior of the network as a whole, allowing us to derive results in a more direct fashion. Our hope is that by further leveraging and framing these problems in a control theoretic context, we can simplify the issue of dealing with adversarial attacks. 

\section{Conclusion} \label{sec:conclusion}
We have, by example, shown how the findings from the field of control theory, and more specifically Lyapunov theory of stability and robustness, are directly applicable and thus related to a new found interest in adversarial attacks. Through this lens, we can more easily define the behavior of networks as a whole, resulting in bounded behavior for all possible inputs. While we do not by any means solve the issue of adversarial attacks in this work, we hope to have effectively illustrated the deep connection between these two fields. 

\newpage

\appendix

\section{Mathematical Reference} \label{sec:prelim}

Our work is based on the Lyapunov theory of stability and robustness for nonlinear dynamical systems, which emerges from the field of control theory.
We recognize many in the machine learning community are not as familiar with this domain of research. 
Here, we give a brief overview of the mathematical principles behind our results and the proposed approach for designing robust DNNs. For a more complete background on stability and robustness of nonlinear systems in the field of control theory, we refer the reader to \cite{khalil2002nonlinear}. 

\begin{figure} 
\centering
\includegraphics[height=3.cm]{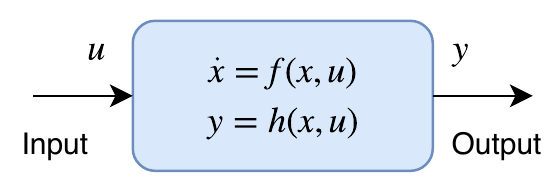}    
\caption{A nonlinear system $H$.}  
\label{Dynsys}
\end{figure}

In our work, we consider the nonlinear system $H$ given in Fig. \ref{Dynsys}, 
\begin{equation*} \label{dynsys}
H:
\begin{cases}
\dot{x}(t)= f(x(t),u(t)) 
&\\y(t)= h(x(t),u(t)),
\end{cases}
\end{equation*}
where  $x(t) \in X \subseteq 
\mathbb{R}^n$, $u(t) \in U \subseteq
\mathbb{R}^m$, and $y(t) \in Y
\subseteq \mathbb{R}^m$ are respectively the state, input and output of the system, and $X$, $U$ and $Y$ are the state, input and output spaces.
\\\\\textbf{Remark 1:} Any layer inside a DNN may be seen as a nonlinear system as described above. For the layer $l$, $u(t)$ has the size of the layer $l-1$ and stands for the input to the layer before the weights and biases are applied. $y(t)$ has the size of layer $l$ and may be seen as the output of the layer $l$ after the activation functions. In this vein, $h$ and $f$ may be thought of as functions which model the state changes ($x(t)$) occurring during the training of the DNN and their relationship to the input and output signals.

\begin{definition} (\cite{zames1966input}) \label{def1}
	System $H$ is instantaneously incrementally finite-gain $L_p$-stable, if for any two inputs $ u_1, u_2\in U$, there exists a positive gain $\gamma$, such that the relation,
	\begin{align*}
	&\|y_{2}-y_{1}\|_{Lp} \leq \gamma\|u_{2}-u_{1}\|_{Lp}.
	\end{align*}
	holds. Here, $\|y_{2}-y_{1}\|_{Lp}$ and $\|u_{2}-u_{1}\|_{Lp}$ represent the $L_p$ Frobenius norm of the signals and $p$ may be any positive number.
\end{definition}
\textbf{Remark 2:} Note that the property defined in Definition \ref{def1} is less restrictive than assuming Lipschitz continuity for a DNN. The Lipschitz property corresponds to replacing the right side of the above equation with a function of input difference i.e., $G(||u_2-u_1||_{L_p})$ which is linear in $||u_2-u_1||_{L_p}$. Further, the above assumption does not place any constraints on the initial conditions of the system DNN. This potentially allows for producing model distributions which have disconnected support \cite{fawzi2018adversarial}.
\begin{definition}  \label{IOFP}
	(\cite{zames1966input})
	System $H$ is considered to be instantaneously Incrementally Output Feedback Passive (IOFP), if it is dissipative with respect to the well-defined supply rate,
	\begin{align*}
\omega(u_2-u_1,y_2-y_1)=(u_2-u_1)^T(y_2-y_1)-\rho (y_2-y_1)^T(y_2-y_1),
	\end{align*}
	for some positive $\rho \in \mathbb{R}$.
\end{definition}
\begin{definition} \label{IIFP}
	(\cite{zames1966input})
	System $H$ is considered to be instantaneously Incrementally Input Feed-Forward Passive (IIFP), if it is dissipative with respect to the well-defined supply rate,
	\begin{align*}
	\omega(u_2-u_1,y_2-y_1)=(u_2-u_1)^T(y_2-y_1)-\nu (u_2-u_1)^T(u_2-u_1),
	\end{align*}
	for some positive $\nu \in \mathbb{R}$.
\end{definition}
\textbf{Remark 3:} A well-defined supply rate function is one that is finite over time and meets certain conditions. System $H$ is dissipative with respect to the well-defined supply rate $\omega(u(t),y(t))$, if there exists a nonnegative storage function $V$ such that $\dot{V}=\omega(u_2-u_1,y_2-y_1)\geq0$. Hence, in order to show that a system is IIFP or IOFP, we need to show that the system's supply rate is greater or equal to zero. For more details on this subject, we refer the readers to \cite{willems1972}. Lastly, the IIFP and IOFP properties of a system have a direct relationship with the system's robustness and stability properties. By proving these properties for each layer of the DNN, we are effectively encouraging the same robust behavior for the entire DNN.
\begin{theorem}
(\cite{khalil2002nonlinear}) If the dynamical system $H$ is Incrementally Output Feedback Passive (IOFP) with $\rho>0$, then it is incrementally finite-gain $L_2$-stable with the gain $\gamma = \frac{1}{\rho}$.
\end{theorem}

\begin{theorem}\label{Th_D} (\cite{arcak2006diagonal})
	A matrix $A \in \mathbb{R}^{n \times n}$ is said to be Lyapunov diagonally stable, if there exists a diagonal matrix $D>0$ such that $DA+A^TD<0.$
\end{theorem}
\begin{theorem} \label{Th_Sec}
	(\cite{arcak2006diagonal}) A matrix of the form:
	\begin{equation*} A = 
	\begin{bmatrix}
	-\alpha_1 & 0 & \dots & 0 & -\beta_N \\
	\beta_1 & -\alpha_2 & \ddots & & 0  \\
	0 & \beta_2 & -\alpha_3  & \ddots & \vdots \\
	\vdots & \ddots & \ddots & \ddots & 0 \\
	0 & \dots & 0 & \beta_{N-1} & -\alpha_N
	\end{bmatrix}, 
	\end{equation*}
	Where $\forall i \in [1, N]$ s.t. $ N > 2, \alpha_i > 0, \beta_i > 0$, is Lyapunov diagonally stable, i.e. it satisfies the relation given in Theorem \ref{Th_D} for some matrix $D>0$,  if and only if the secant criterion,
	\begin{equation*}
	\frac{ \beta_1 \times  \beta_2\times \dots \times  \beta_N}{\alpha_1 \times\alpha_2\times\dots\times\alpha_N}<\sec\left(\frac{\pi}{N}\right)^N=\frac{1}{\cos\left(\frac{\pi}{N}\right)^N}
	\end{equation*}
	holds.
\end{theorem}
\textbf{Remark 4:} It is important to note that the properties given in above theorems will be utilized in our proofs 
to show stability and robustness for a cascade of layers in a DNN.

\section{Empirical Experiments} \label{sec:experiments}

The primary purpose of our paper is to show, by example, the connection of control theory to adversarial attacks. Having bounded the deviation of a network's activation's given a constant perturbation to the input, we now show this holds empirically and can be adapted into a regularize with little work. Because our adversary is constraint to a single constant perturbation, this is of little practical importance. It however is indicative of how more involved applications of control theory may be adapted into usable defenses, and empirically confirms our proof holds. 

The regularization term defined in \eqref{eq:regularization_term} directly from the proof. It is composed of layer dependent penalties, each with two components. A "constant" term that is determined by the values of the hyper parameters ($\nu_l$, the Leaky ReLU slope $a$, and hidden layer size $n_l$) which are defined before training starts.  Subtracted from this is a "weight" term, which is simply the sum of all weight coefficients for the hidden layer. The penalty simply encourages the sum of weights in layer $l$ to be larger than the layer-wise constant $n_{l-1} \cdot v_l \cdot a^{-1}$. 

\begin{equation}\label{eq:regularization_term}
    \sum_{l=1}^N \max \left( \underbrace{n_{l-1} \cdot \nu_l \cdot  a^{-1}}_{\text{Constant term}} - \underbrace{\sum_{i=1}^{n_l} \sum_{j=1}^{n_{l-1}} w^l_{ij}}_{\text{Weight term}} ,  0 \right)
\end{equation}

\subsection{Dataset Details} \label{sec:data_details}

Experiments were run on the following regression datasets. Prior to use, each dataset was split into training, validation, and test sets. The independent variables for each dataset were normalized using training set statistics, and principal component analysis was used to reduce the dimensionality of each dataset to 10. The target variable for each dataset was also scaled to the $[0,1]$ range. All data sets were obtained from the UCI Machine Learning repository. 


\textit{Boston Housing:} the first dataset we evaluate on is the Boston house price data of \textcite{Harrison1978HedonicAir}, the target variable is the median value in thousands of dollars of owner-occupied homes in the area of Boston, Massachusetts. 

\textit{Communities and Crime:} the second data set we evaluate on is the Communities and Crime Unnormalized data set \cite{REDMOND2002660}. The number of murders in 1995 is the target variable, and variables include potential factors such as percent of housing occupied, per capita income, and police operating budget. Independent variables from the original data set that contained missing values were dropped.

\textit{Relative Location of CT Slices on Axial Axis:} the third data set we evaluate on is the Relative Location of CT Slices on Axial Axis data set \cite{Graf2011}. The data consists of a set of 53500 CT images from 74 different patients where each CT slice is described by two histograms in polar space. The histograms describe the location of bone structures in the image and the location of air inclusions inside of the body. The independent variables consist of the information contained in the two histograms, and the target variable is the relative location of an image on the axial axis.


\textit{Malware:} The fourth data set we evaluate on is the Dynamic Features of VirusShare Executables data set from \textcite{Huynh2017} which contains the dynamic features of executables collected by VirusShare between November 2010 and July 2014. The target variable is a risk score between 0 and 1. This data set is an intrinsically interesting use case as malware authors are an active real-life adversary. 


\textit{Condition Based Maintenance:} The fifth data set we evaluate our approach on is the Condition Based Maintenance of Naval Propulsion Plants data set consists of results from a numerical simulator of a naval vessel characterized by a gas turbine propulsion plant \cite{Coraddu2014}. 
This
data set has two target variables, the gas turbine's compressor decay state coefficient and the gas turbine's turbine decay state coefficient. As such we will treat this as two different regression data sets that use the same feature set.

\subsection{Network Architecture, Training, and Attack Settings} \label{sec:experiment_design}

The network architectures in all experiments consist of an input layer, 2, 6, or 12 hidden layers of size equal to the input layer size, and a single node output layer. Leaky rectified linear units were used as hidden layer activation functions with the negative slope set to $a = 0.5$, and Adam was used as the optimization algorithm \cite{Kingma2015}. The regularization term from \eqref{eq:regularization_term} was re-scaled to receive a weight (i.e., magnitude) equal to the mean squared error in the loss function. This was to avoid numerical issues in training. In all cases, the hyper-parameter $\nu_l$ is set to $1$, i.e. $\nu_l = 1.0$ for $l=1,...,N$. While training with gradient descent in this fashion does not guarantee that the conditions will be met for $\nu_l = 1$ for $l=1,...,N$, in practice the results are close, with more details shown in appendix \autoref{sec:nu_learned_is_close}. The adversarial attack follows Definition \ref{attack} with $\epsilon = 0.5$. Because of the constraint $\Delta^{(1)}_i = \Delta^{(1)}_j$, $\forall i, j$, our threat model results in an adversary with a single degree of freedom. We are aware this is a weaker threat model, but our focus is to show the connections between the Lyapunov theory and the domain of adversarial attacks. It also allows us to use Hill climbing to find the optimal perturbation vector within the $\epsilon$-bounded $L_p$-ball.

\subsection{Results and Discussion}

Bounds for each dataset and network depth combination are computed by finding the largest values of $\nu_l$ for each layer satisfying Equation \ref{key} and the learned weights parameters. 


\begin{figure*}[!htb]
  \centerline{\includegraphics[width=\textwidth]{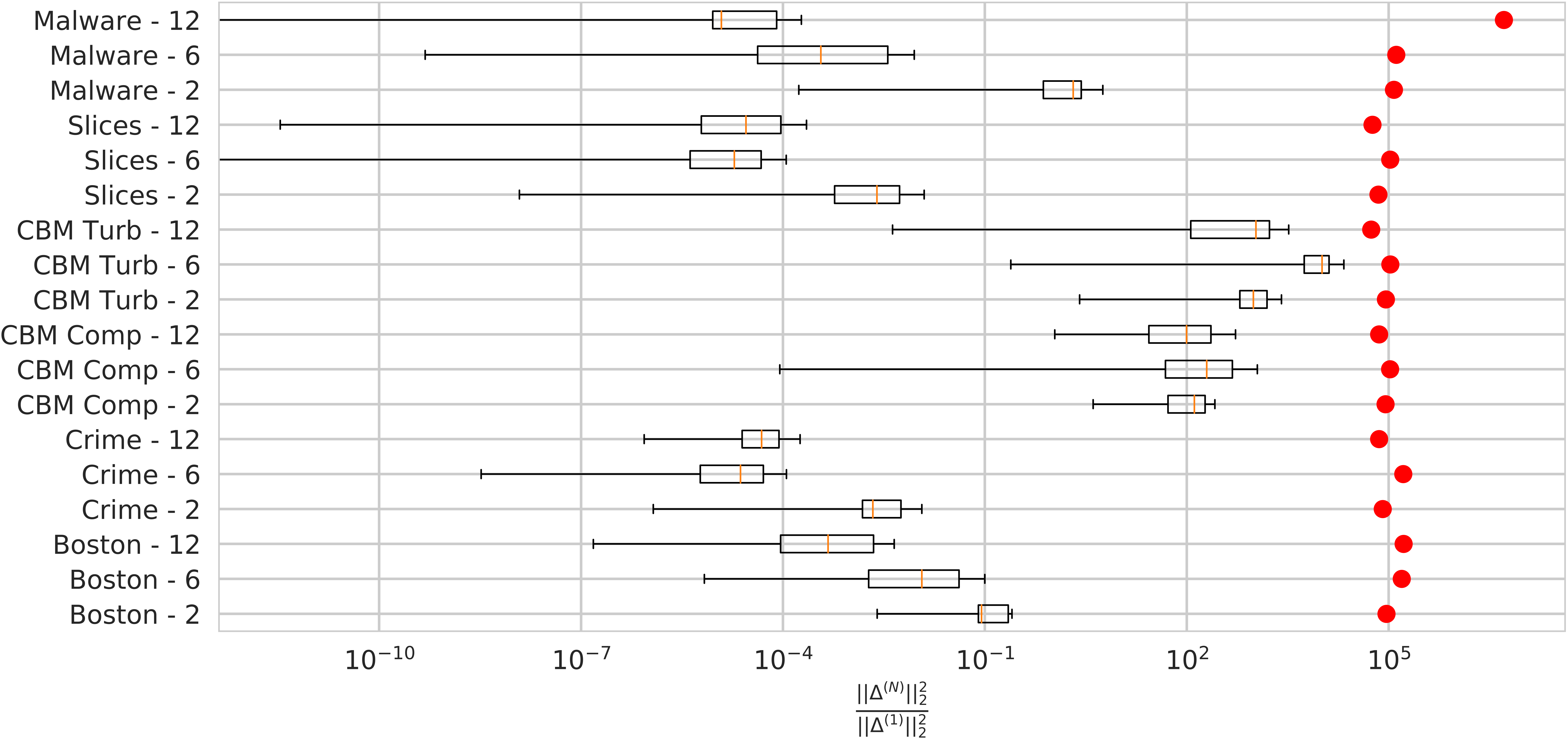}}
  \caption{Box plot describing the distribution of $\frac{||\Delta^{(N)}||^2_2}{||\Delta^{(1)}||^2_2}$ for each dataset and network depth combination. Red dots represent the upper bound on the ratio, 
based on \autoref{eq:final_bound_general}. 
  }
  \label{fig:ratios}
\end{figure*}

Figure \ref{fig:ratios} shows a box plot describing the distribution of $\frac{||\Delta^{(N)}||^2_2}{||\Delta^{(1)}||^2_2}$ for each dataset and network depth combination. Red dots represent the upper bound on the ratio, computed as $\frac{\beta}{2(\epsilon-\rho-{\frac{1}{2\beta}})}$. This plot illustrates the fact that no bound violations occurred in any of our experiments. For all test data points, datasets, and network depths, $\frac{||\Delta^{(N)}||^2_2}{||\Delta^{(1)}||^2_2}$ is lower than the upper bound.

For many of the datasets, there is a wide gap between the worst observed perturbations resulting from an attack and the upper bound on the perturbations. The presence of large gaps suggests that there could be a difference between the bound on the universe of \textit{all possible} data (what our theory provides) and the space occupied by \textit{observed} data. This suggests room for tighter bounds by looking at the behavior of the network within a limited input space defined by the data. 
However, in some cases such as on the CBM Turbine dataset with 6 hidden layers, the gap is much smaller, demonstrating that bounding the worst case scenario does have practical value as it is possible in practice to get very close to the worst case.

\subsection{Learn $\nu_l$ are close to the desired value. } \label{sec:nu_learned_is_close}

\begin{figure*}
  \centerline{\includegraphics[width=\textwidth]{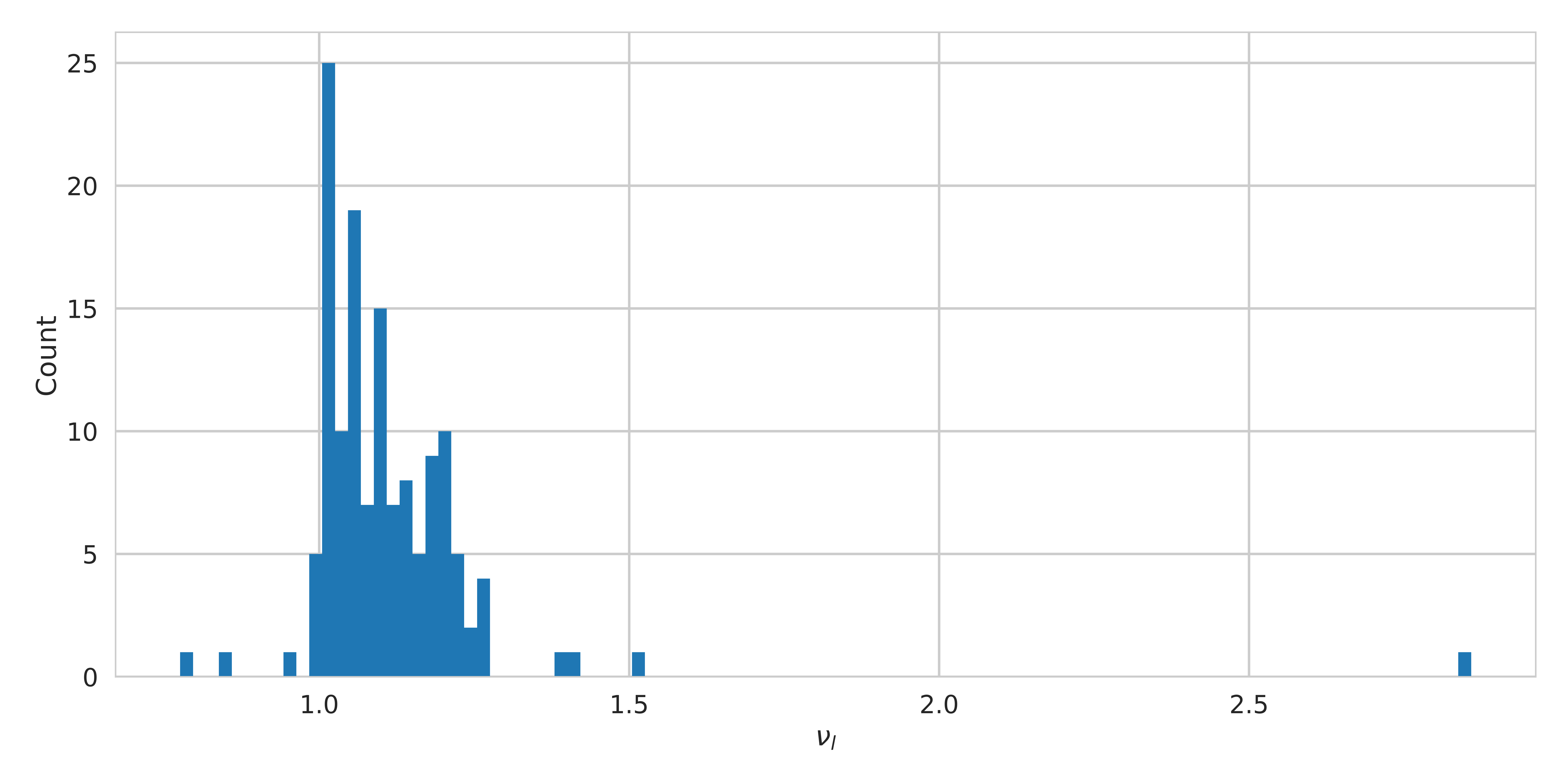}}
  \caption{Histogram describing the distribution of $\nu_l$ computed from the trained networks combined across all datasets and network depths using Equation \ref{key}.}
  \label{fig:nus}
\end{figure*}

Figure \ref{fig:nus} contains a histogram illustrating the distribution of $\nu_l$ computed from the trained networks combined across all datasets and network depths using Equation \ref{key}. As described in Section \ref{sec:experiments}, in the regularization term the desired $\nu_l$ were set to $\nu_l = 1.0$ for all layers in all networks. This figure shows that the resulting $\nu_l$ are almost all equal or slightly greater than $1.0$, noting that larger $\nu_l$ result in more resilient networks. This demonstrates the practical effectiveness of the regularization term defined in \eqref{eq:regularization_term} at obtaining networks with a specified desired resilience.

It would be possible to force the chosen value of $\nu_l=1$ to occur in the network by using a projection step after every gradient update. Not only is this more computationally demanding, but we find it makes learning a network with comparable MSE more difficult. Because we see that the learn value of $\nu_l$ is almost always closer to the desired value, we prefer to train in this relaxed fashion, and can be seen as a way of allowing the network flexibility to reduce the bound in order to obtain a useful model. This is reasonable in our opinion, since a model with degenerate performance in all cases is intrinsically never useful.


\bibliographystyle{ACM-Reference-Format}
\bibliography{Mendeley,arashbib}

\end{document}